\documentclass[twocolumn,trackchanges]{aastex631}

\usepackage{amsmath}
\usepackage{comment}
\usepackage{xcolor}

\shorttitle{Reflected light coronagraphy simulations}
\shortauthors{Kammerer et al.}
\graphicspath{{./}{figures/}}

\begin{document}

\title{Simulating reflected light coronagraphy of Earth-like exoplanets with a large IR/O/UV space telescope: impact and calibration of smooth exozodiacal dust}

\author[0000-0003-2769-0438]{Jens Kammerer}
\affiliation{Space Telescope Science Institute\\
3700 San Martin Drive\\
Baltimore, MD 21218, USA}

\author{Christopher C. Stark}
\affiliation{NASA Goddard Space Flight Center\\
Greenbelt, MD 20771, USA}

\author{Kevin J. Ludwick}
\affiliation{Center for Applied Optics\\
University of Alabama in Huntsville\\
301 Sparkman Dr NW\\
Huntsville, AL 35805, USA}

\author{Roser Juanola-Parramon}
\affiliation{NASA Goddard Space Flight Center\\
Greenbelt, MD 20771, USA}
\affiliation{University of Maryland Baltimore County\\
Baltimore, MD 21250, USA}

\author{Bijan Nemati}
\affiliation{Tellus1 Scientific LLC\\
Madison, AL 35758, USA}







\begin{abstract}

Observing Earth-like exoplanets orbiting within the habitable zone of Sun-like stars and studying their atmospheres in reflected starlight requires contrasts of $\sim1\mathrm{e}{-10}$ in the visible. At such high contrast, starlight reflected by exozodiacal dust is expected to be a significant source of contamination. Here, we present high-fidelity simulations of coronagraphic observations of a synthetic Solar System located at a distance of 10~pc and observed with a 12~m and an 8~m circumscribed aperture diameter space telescope operating at 500~nm wavelength. We explore different techniques to subtract the exozodi and stellar speckles from the simulated images in the face-on, the 30~deg inclined, and the 60~deg inclined case and quantify the remaining systematic noise as a function of the exozodiacal dust level of the system. We find that in the face-on case, the exozodi can be subtracted down to the photon noise limit for exozodi levels up to $\sim1000$~zodi using a simple toy model for the exozodiacal disk, whereas in the 60~deg inclined case this only works up to $\sim50$~zodi. We also investigate the impact of larger wavefront errors and larger system distance, finding that while the former have no significant impact, the latter has a strong (negative) impact. Ultimately, we derive a penalty factor as a function of the exozodi level and system inclination that should be considered in exoplanet yield studies as a realistic estimate for the excess systematic noise from the exozodi.

\end{abstract}



\section{Introduction}
\label{sec:introduction}

The direct detection and characterization of Earth-like exoplanets orbiting within the habitable zone of Sun-like stars has recently been recommended by the Astro2020 Decadal Survey on Astronomy and Astrophysics to be one of the key science drivers for the next NASA flagship mission: a large $\sim6$~m infrared/optical/ultraviolet (IR/O/UV) space telescope. Such a mission could, for the first time, reveal the presence of surface oceans and search for so-called biomarkers in the atmospheres of other worlds \citep{lustig-yaeger2018,luvoir2019}. However, one hurdle to overcome on this endeavor are small dust grains believed to exist in virtually all (exo-)planetary systems which reflect a small fraction of the host star light and can bury the signal of a potential exo-Earth \citep{roberge2012}. In our own Solar System, this dust is referred to as zodiacal dust and in exoplanetary systems, it is often called exozodiacal dust.

Previous studies have shown that the exo-Earth candidate yield of a future IR/O/UV space telescope, that is the yield of Earth-like exoplanets orbiting within the habitable zone of their host star, is a strong function of the telescope aperture size \citep{stark2014,stark2015}. For a $\sim6$~m-class telescope and an average exozodi level of 3~zodi, \citet{stark2015} find an expected exo-Earth candidate yield of $\sim10$ if the occurrence rate $\eta_\oplus$ of such planets is approximately 10\%. While these same studies suggest that the yield is a weak function of the average exozodi level, it is important to note that these studies assume that the exozodi can be subtracted down to the photon noise limit, an assumption that has not yet been well studied. The validity of this assumption is not trivial given the complex morphologies (e.g., ring structures, radial color gradients, azimuthal variations due to the scattering phase function) that exozodiacal dust disks are expected to exhibit \citep[e.g.][]{jackson1989,hughes2018}. The presence of multiple small inner planets can further complicate the disk morphology due to dynamical interactions between the planets and the dust grains \citep[e.g.][]{dermott1995}. For a future space-based mid-infrared nulling interferometer searching for habitable exoplanets \citep{cockell2009,quanz2021_exa}, \citet{defrere2010} have investigated the impact of exozodiacal dust on the detectability of an Earth twin at 15~pc in the presence of geometrical offsets of the exozodiacal dust cloud center from the host star or planet-induced resonant structures. They found that geometrical offsets of $<0.5$~au are acceptable for exozodi levels $\lesssim50$~zodi and planet-induced resonant structures become an issue at $10~\text{\textmu m}$ for exozodi levels $\gtrsim15$~zodi at 0~deg inclination and $\gtrsim7$~zodi at 60~deg inclination.

Here, we study our ability to subtract the exozodi at visible wavelengths in the $1\mathrm{e}{-10}$ contrast regime. To do so, we present simulations of coronagraphic observations of a synthetic Solar System generated with the \texttt{exoVista} package\footnote{\url{https://starkspace.com/}} \citep{stark2022}. In our simulations, we explore different wavefront errors, system distances, and telescope aperture sizes and quantitatively assess the contamination arising from the exozodi. Depending on the amount of dust and the distance of the system, clever observing and post-processing techniques are required to reveal the signatures of potentially habitable planets. Here, we simulate roll subtraction and (ideal) reference star subtraction techniques and calibrate residual flux from the exozodiacal disk using a simple toy model for the disk. Here, ideal means that the simulated PSF reference star has exactly the same properties as the science target, so that the only difference between the two observations are the photon noise and the changes in the speckle field due to wavefront drifts. Ultimately, our simulation tools and data reduction techniques can be applied to a full synthetic population of nearby exoplanetary systems. This would enable modeling a realistic execution of an observing sequence of a future IR/O/UV space telescope and studying real-time decision making metrics aiming at maximizing the yield of potentially habitable exoplanets.

\section{Methods}
\label{sec:methods}

To generate the simulated observations, we convolve scene images from \texttt{exoVista} with simulated spatially-dependent point-spread functions (PSFs) of a high-contrast apodized pupil Lyot coronagraph \citep[APLC,][]{aime2002,soummer2005} and a high-angular resolution vector vortex charge 6 coronagraph \citep[VC6,][]{mawet2010}. Then, we add photon and detector noise using the \texttt{emccd\_detect} package\footnote{\url{https://github.com/wfirst-cgi/emccd_detect}} \citep{nemati2020} to the noiseless images. Our simulations include stellar speckles, exoplanets, and exozodiacal dust and cover a range of exozodi levels from 1 to 1000~zodi in the face-on, the 30~deg inclined, and the 60~deg inclined case. After performing either a roll subtraction or an (ideal) reference star subtraction, we fit a simple toy model consisting of a fifth order polynomial in radius and a tenth order polynomial in azimuth to the residual exozodiacal disk signal. After further subtracting the best fit disk model, we then measure the residual noise in the simulated images and compare it to the theoretically expected photon noise limit in order to asses whether the previously made simplifying assumption, namely that the exozodi can be subtracted down to the photon noise limit, is valid or not.

\subsection{Simulating coronagraphic observations}
\label{sec:simulating_coronagraphic_observations}

\subsubsection{Synthetic Solar System}
\label{sec:synthetic_solar_system}

The first step in our simulation process is to generate astrophysical scenes of a synthetic Solar System located at a distance of 10~pc using the \texttt{exoVista} package \citep{stark2022}. For this purpose, we use the \texttt{generate\_solarsystem} routine provided by \texttt{exoVista} which simulates the Solar System planets on temporally evolving orbits, including their varying reflected light spectra, as well as a static multi-component exozodiacal dust disk. For simplicity, we ignore the cold component of the Solar System dust complex and only include the exozodiacal component. We note that \texttt{exoVista} only models symmetric exozodi meaning that we are only considering smooth, symmetric dust disks without additional structure here.

For investigating the impact of the system distance, we also simulate a Solar System analog at a distance of 15~pc. To factor out the impact of decreasing angular resolution, we increase the orbits of the planets and the radius of the exozodiacal dust disk by 50\% and replace the Sun with a 125\% more luminous F-type star.

\begin{figure*}
\centering
\includegraphics[width=\textwidth]{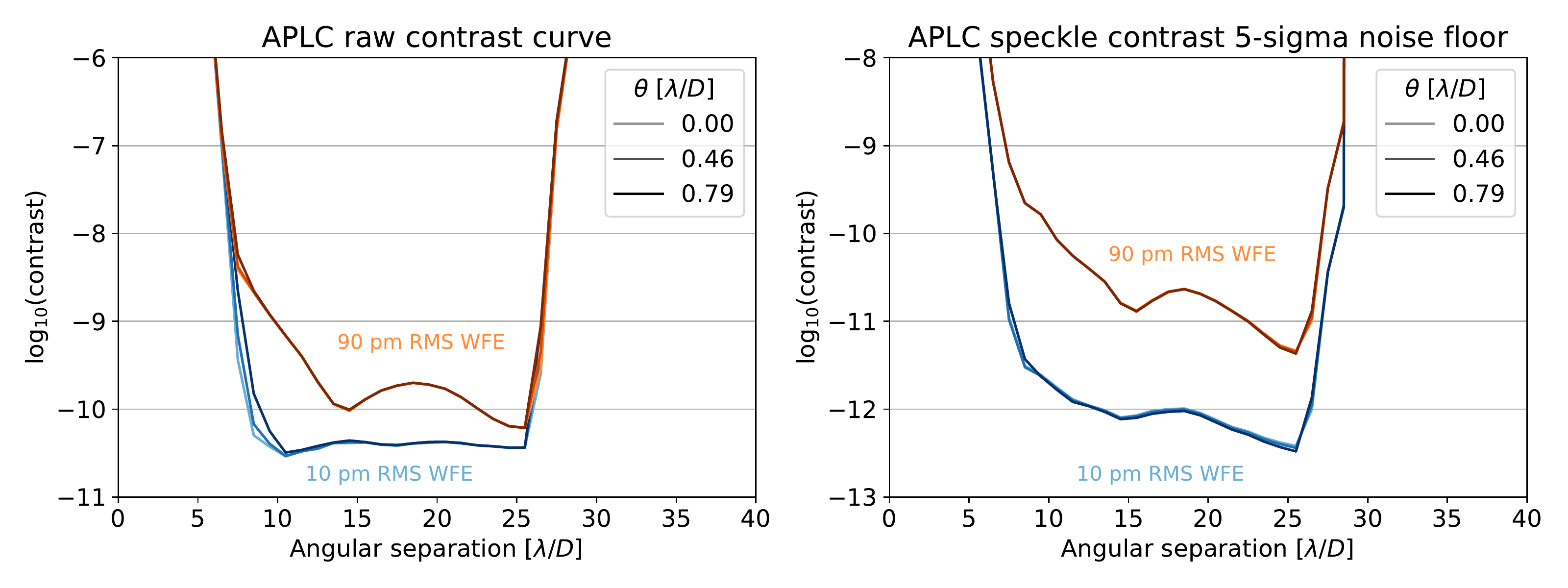}
\includegraphics[width=\textwidth]{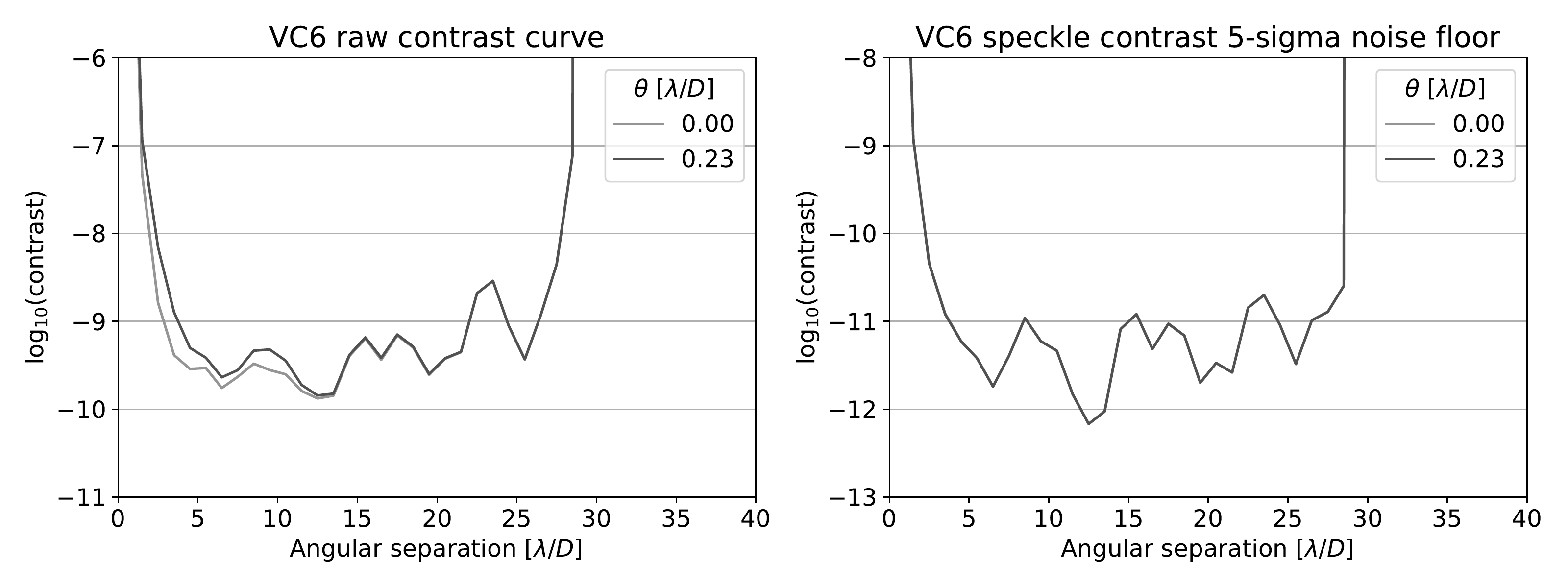}
\caption{Raw contrast (left panels) and speckle noise floor after subtracting a PSF reference star (right panels) of the LUVOIR-A apodized pupil Lyot coronagraph (APLC, top panels) and the LUVOIR-B vector vortex charge 6 coronagraph (VC6, bottom panels) used in this work to simulate coronagraphic observations of exoplanetary systems. The different shadings represent different stellar angular diameters given in units of $\lambda/D$. In the top panels, the blue curves correspond to a model assuming a wavefront error of 10~pm RMS and the orange curves correspond to a model assuming a wavefront error of 90~pm RMS.}
\label{fig:aplc}
\end{figure*}

\begin{figure*}
\centering
\includegraphics[width=\textwidth]{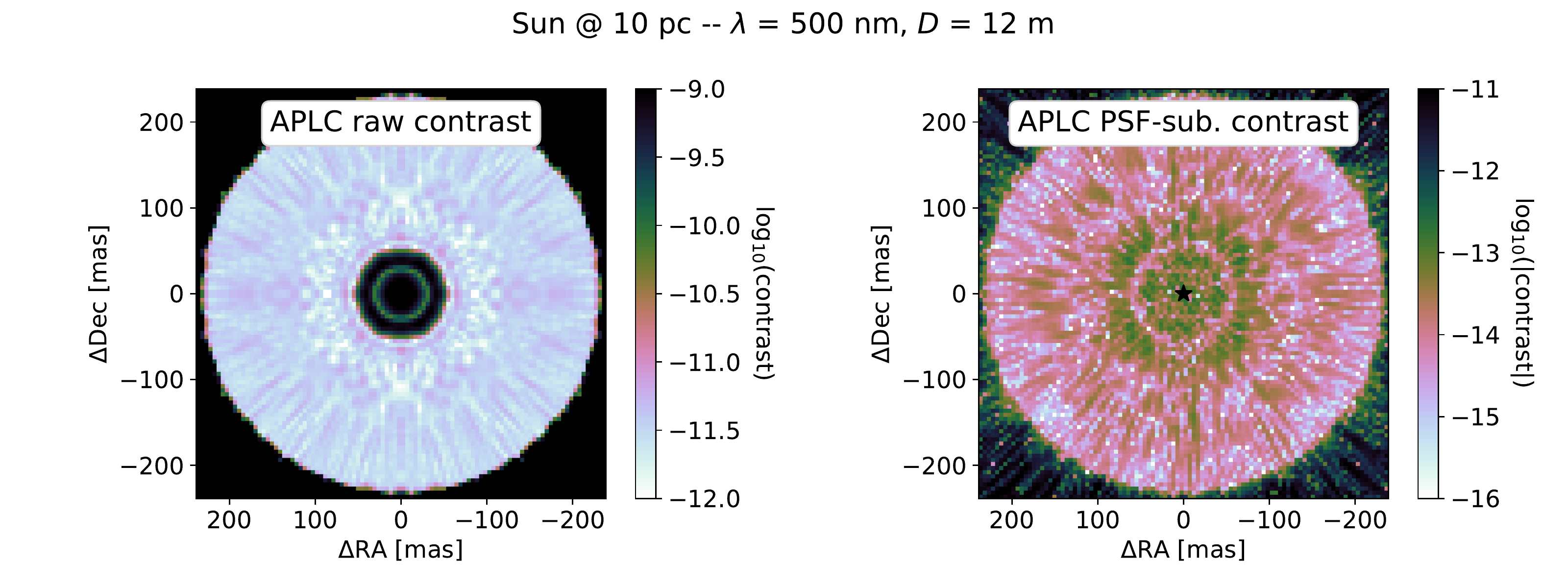}
\includegraphics[width=\textwidth]{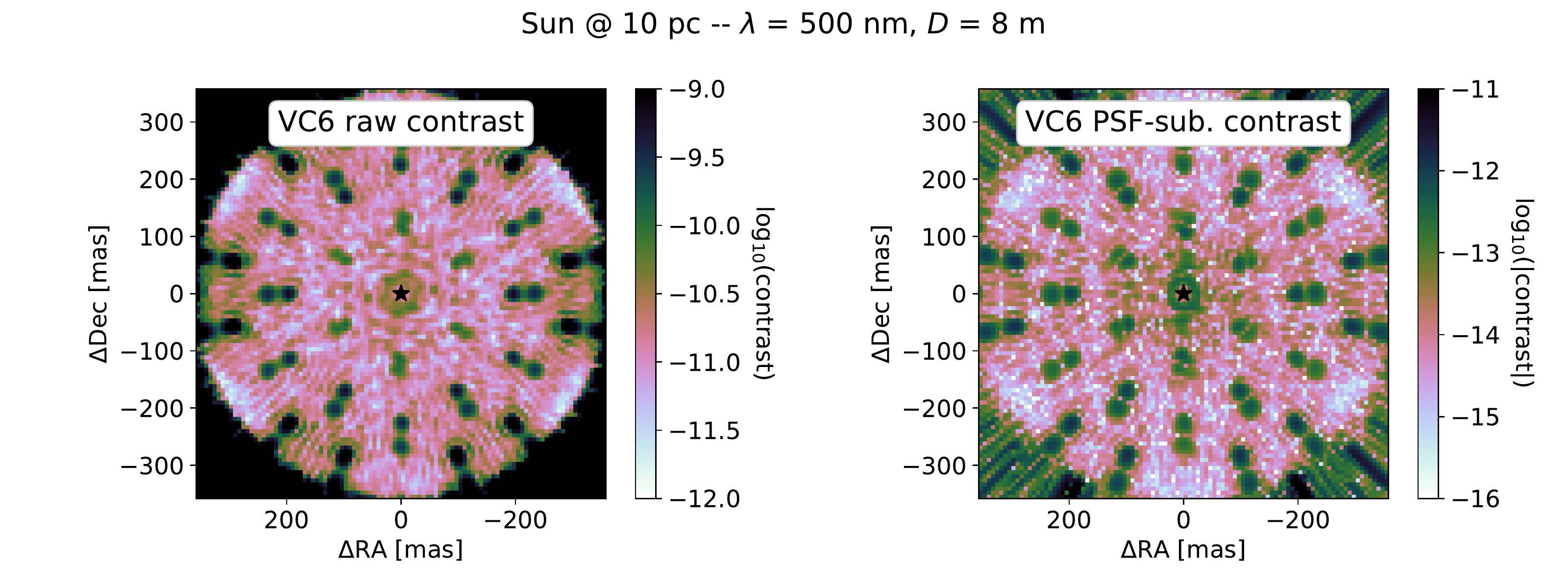}
\caption{Same as Figure~\ref{fig:aplc} but showing the full 2D speckle fields instead of the azimuthally-averaged 5--$\sigma$ contrast curves. Here, the baseline (10~pm RMS WFE) case is shown for the APLC in the top panels. The images are evaluated for observations of a Solar analog at a distance of 10~pc and an observing wavelength of 500~nm.}
\label{fig:speckles}
\end{figure*}

\subsubsection{Coronagraph models}
\label{sec:coronagraph_models}

\begin{figure*}
\centering
\includegraphics[width=\textwidth]{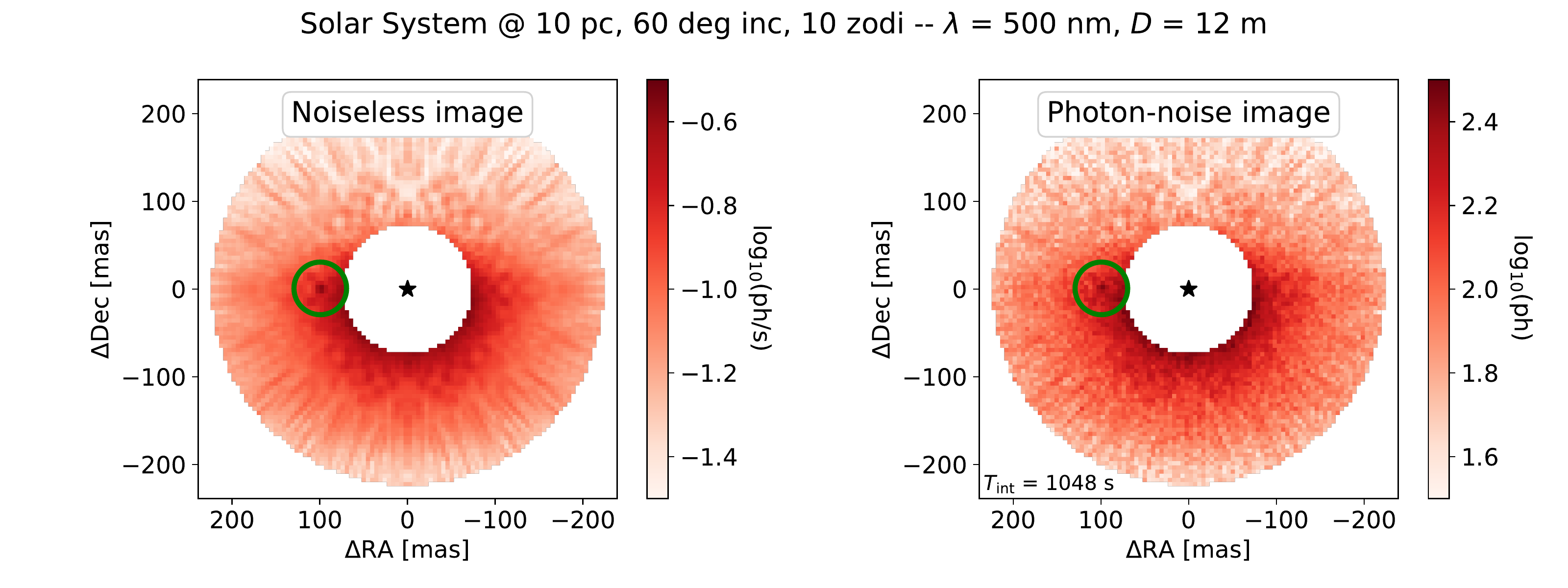}
\caption{Noiseless coronagraphic scene image obtained by convolving the \texttt{exoVista} scene with the LUVOIR-A APLC PSF (left panel) and the same image with photon noise added and downsampled to a detector pixel scale of $4.77$~mas (right panel). The image shows the contribution of the companions, the exozodiacal dust disk, and stellar speckles behind the coronagraph for a 60~deg inclined Solar System with 10~zodi of dust and at a distance of 10~pc. The position of the host star (here the Sun) is indicated by a black star and the position of the planetary-mass companion (here the Earth) is indicated by a green circle. It can be seen that one obtains less than one photon per second from an Earth-twin at a distance of 10~pc and a wavelength of 500~nm observed through an 18\% bandpass (assuming a total system throughput of $\sim31.78\%$).}
\label{fig:exovista}
\end{figure*}

The next step in our simulation process consists of convolving the astrophysical scenes from \texttt{exoVista} with LUVOIR-A and LUVOIR-B coronagraph models. For most of our simulations, we adopt the same coronagraph model used in the \emph{LUVOIR} study report for the majority of exo-Earth candidate detections \citep{luvoir2019} which resembles an 18\% bandwidth APLC achieving a raw contrast of $<1\mathrm{e}{-10}$ beyond $\sim8~\lambda/D$ separation for sufficiently small stellar angular diameters ($\lesssim0.5~\lambda/D$, where $\lambda$ is the observing wavelength and $D$ is the telescope primary mirror diameter, see top left panel of Figure~\ref{fig:aplc}). However, to investigate the impact of telescope aperture size, we also present simulations with a VC6 coronagraph achieving a speckle noise floor of $<1\mathrm{e}{-10}$ already beyond $\sim2.5~\lambda/D$ separation (see bottom right panel of Figure~\ref{fig:aplc}). The coronagraphic PSFs are 239 by 239~pixel images with a pixel scale of $0.25~\lambda/D$ so that they can easily be scaled to an arbitrary observing wavelength and telescope aperture size. The core throughput of the APLC is $\sim32\%$, that of the VC6 is $\sim36\%$, and the instrument throughput is assumed to be 100\%. While we consider temporal wavefront drifts of 10, 30, and 90~pm RMS, we do not consider static wavefront or target acquisition errors since we aim to study exozodi subtraction methods in the $1\mathrm{e}{-10}$ contrast regime and not all possible degradations of the coronagraph performance.

For simulating the PSF of the host star, we generate two sets of on-axis PSFs, each featuring a different speckle field. The speckles vary by less than 1\% of the raw contrast of the APLC, that is approximately $\sim1\mathrm{e}{-13}$ (see top right panel of Figure~\ref{fig:aplc}). Each set consists of multiple PSFs simulated for different stellar angular diameters and the two different sets can be used to generate images with two different telescope roll angles or of a science and a reference target observation. The speckle field of these two sets of on-axis PSFs is the result of the propagation of time-dependent dynamic aberrations. These are introduced as time series in which the optical path difference (OPD) error maps present at the entrance pupil of the coronagraph vary as a function of time (during 20~seconds represented with 8000~two-dimensional OPD maps). This time series was generated by Lockheed Martin via an integrated model of the telescope and spacecraft structural dynamics, and includes the rigid body motion of the primary mirror segments, the dynamic interaction of flexible structures, and the disturbances from the pointing control system. For reference, the full 2D speckle fields before and after subtracting a reference PSF are shown for the APLC with a temporal wavefront drift of 10~pm RMS and the VC6 in Figure~\ref{fig:speckles}.

For simulating the PSFs of companions (such as planets) and the exozodiacal dust disk, we generate a set of off-axis PSFs, computed for different radial offsets and position angles. For the companions, we simply interpolate this grid of off-axis PSFs for the radial offset and position angle of the respective companion. To convolve the exozodiacal dust disk, we pre-compute a data cube of off-axis PSFs for each pixel on a 239 by 239~pixel grid with a pixel scale of $0.25~\lambda/D$ and scale the scene image of the disk from \texttt{exoVista} to that same pixel scale. Then, we convolve the scene image with the data cube of spatially-dependent PSFs. We note that this is a memory-intense convolution of a $239^2$~pixel image with a $239^4$~pixel data cube of PSFs. The four-dimensional data cube of PSFs (there is a two-dimensional offset PSF for each pixel in the two-dimensional field-of-view) is $\sim26$~GB in size and takes roughly three minutes to compute on a 2020 MacBook Pro with 32~GB RAM and Intel Core i7 Quad-Core CPU. Calculating a new grid of PSFs for each wavelength or astrophysical scene would therefore dominate the runtime of our numerical simulations. To avoid this, we adopt a pixel scale in units of $\lambda/D$ so that the PSF grid can be applied to an arbitrary observing wavelength and telescope aperture size. We then interpolate the astrophysical scene to the resolution of our PSF data cube and perform the convolution via matrix multiplication. With a computer where the entire data cube of PSFs ($\sim26$~GB) fits into the RAM, we find that the \texttt{NumPy} \texttt{tensordot} function\footnote{\url{https://numpy.org/doc/stable/reference/generated/numpy.tensordot.html}} delivers the best performance. While the first convolution operation is equally slow as with other \texttt{NumPy} functions ($\sim12$~s in our case), subsequent convolution operations are more than one order of magnitude faster than the first one ($\sim0.6$~s in our case) since the data cube of PSFs is already stored in the correctly aligned format for multiplication with the BLAS library\footnote{\url{http://www.netlib.org/blas/}} in the RAM. This is especially useful when synthesizing images of multiple exoplanetary systems, multiple epochs, or multiple wavelengths at once.

\subsubsection{Photon and detector noise}
\label{sec:photon_and_detector_noise}

\begin{figure*}
\centering
\includegraphics[trim={0 3cm 0 1cm},clip,width=\textwidth]{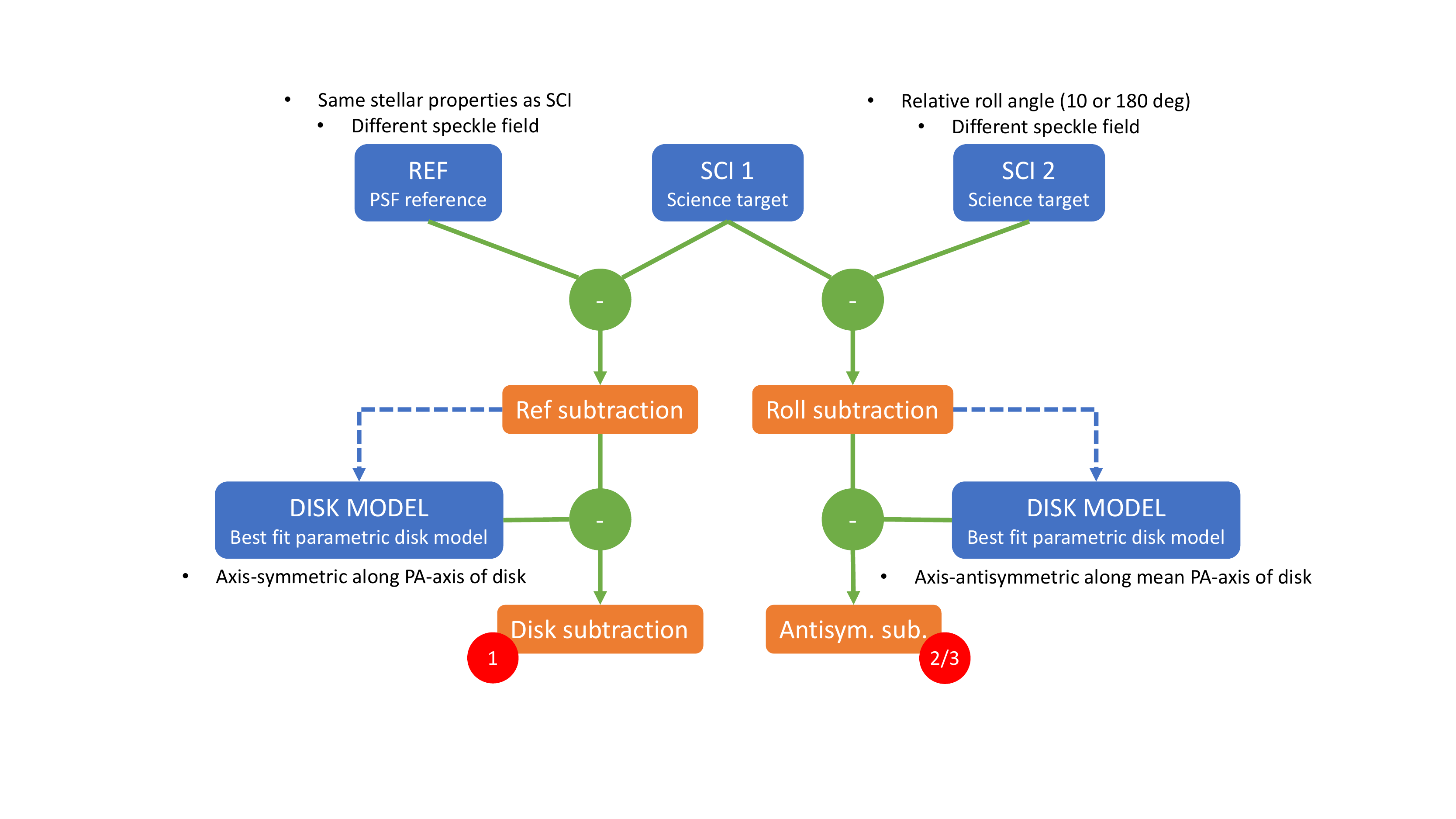}
\caption{Schematic illustrating the different PSF and disk subtraction methods used in this work. We start by simulating coronagraphic images of a PSF reference (REF) and the science target (SCI 1) as well as the science target with a roll angle of 10/20 and 180~deg (SCI 2). Then, we perform reference star or roll subtraction to calibrate stellar speckles and subsequently subtract a best fit toy model for the residual exozodiacal disk flux from the images. In the resulting science products (disk and antisymmetric subtraction), we then measure the residual noise and compare it to the theoretical photon and detector noise limit.}
\label{fig:psf_and_disk_subtraction_methods}
\end{figure*}

\begin{table}
\centering
\caption{Detector parameters adopted for the noise modeling of the EMCCD detector.}
\begin{tabular}{lll}
Parameter & Value & Unit\\
\hline
Electron-multiplying gain & 5000 & e-/photon\\
Image area full well capacity & 60000 & e-\\
Serial register full well capacity & 100000 & e-\\
Dark current & $3\mathrm{e}{-5}$ & e-/s\\
Clock-induced charge & $1.3\mathrm{e}{-3}$ & e-/frame\\
Read noise & 0 & e-/frame\\
Bias offset & 10000 & e-\\
Quantum efficiency & 0.9 & --\\
Cosmic ray rate & 0 & hits/cm$^2$/s\\
Pixel pitch & $1.3\mathrm{e}{-5}$ & m\\
Detector gain & 1 & e-/ADU\\
\hline
\end{tabular}
\label{tab:detector_parameters}
\end{table}

In this work, we aim to use simple methods to subtract exozodi from simulated images of a future IR/O/UV space telescope and to investigate whether the assumption made in previous yield estimates, namely that the exozodi can be subtracted down to the photon noise limit, is valid or not. In the photon noise limit, the integration time $T_\text{int}$ required to detect a faint companion at a given signal-to-noise ratio SNR is given as
\begin{equation}
    \label{eqn:tint}
    T_\text{int} = \text{SNR}^2\left(\frac{\text{CR}_\text{plan}+2\text{CR}_\text{back}}{\text{CR}_\text{plan}^2}\right),
\end{equation}
where $\text{CR}_\text{plan}$ is the count rate of the planet that one aims to detect (here, this is an Earth twin at quadrature), and $\text{CR}_\text{back}$ includes all relevant background sources such as the residual stellar speckles, the exozodiacal dust disk, and detector noise \citep{stark2019}. Here, we directly measure the count rate of the residual stellar speckles, the exozodiacal dust disk, and the planets in the simulated noiseless images by integrating over a photometric aperture with a radius of $0.8~\lambda/D$ (where $D$ is the circumscribed mirror diameter) placed at the position of an Earth-twin at quadrature. The detector noise is estimated according to \citet{stark2019} as
\begin{equation}
    \text{CR}_\text{detn} = N_\text{pix}\left(d+\frac{r^2+c}{T_\text{frame}}\right),
\end{equation}
where $N_\text{pix}$ is the number of detector pixels within the $0.8~\lambda/D$ photometric aperture, $d$ is the dark current, $r$ is the read noise, $c$ is the clock-induced charge, and $T_\text{frame}$ is the detector frame time. Ultimately, we will then compare the noise predicted by the aforementioned equations with the true noise measured in the simulated and reduced images to assess whether the exozodi was subtracted down to the photon noise limit or not. Table~\ref{tab:detector_parameters} presents the detector parameters adopted in this work.

After convolving the astrophysical scenes from \texttt{exoVista} with the coronagraphic PSFs in the previous Section, we now scale the images to a constant detector pixel scale and add photon and detector noise. We assume that the detector is Nyquist-sampled at $\lambda_\text{Nyquist} = 500$~nm, yielding a pixel scale $s$ of
\begin{equation}
    s = 0.5\frac{\lambda_\text{Nyquist}}{0.9D} = 4.77~\text{mas}
\end{equation}
for our descoped version of LUVOIR-A with $D = 12$~m, where we are here considering the inscribed mirror diameter which is $0.9D = 10.8$~m. Given that the outer working angle (OWA) of the APLC coronagraph is $\sim26~\lambda/D = 223$~mas, we choose an image size of 100 by 100 pixels yielding a field-of-view of 477~mas sufficient to fit the entire coronagraphic scene image. We also convert the image units from Jansky per pixel to photons per pixel. For LUVOIR-B with $D = 8$~m, the pixel scale is twice as large ($9.55$~mas) and we adopt an OWA of $13~\lambda/D$ since the exozodiacal disk scene from \texttt{exoVista} only extends to $\pm250$~mas.

For the purpose of directly observing Earth-like exoplanets around Sun-like stars at distances of $\gtrsim10$~pc, it is essential to have a detector that can count each individual photon received from such a planet since the expected count rate is quite low (often less than one photon per second). There are multiple possible detector solutions to enable photon counting with low noise properties. Our intent here is not to recommend one such detector technology, but rather to explore the potential impact of detector noise on exozodi background subtraction. As such, we choose to model an Electron-Multiplying Charge-Coupled Device (EMCCD), which is being used for the \emph{Roman} Coronagraphic Instrument \citep[CGI,][]{kasdin2020} and was also discussed in the \emph{LUVOIR} study report \citep{luvoir2019}. With an EMCCD, the detector is read out on a high cadence to ensure that each individual photon received from a potential planet is counted and many individual reads are combined into a single exposure. Hence, our observations are characterized by a detector frame time $T_\text{frame}$ and a detector integration time $T_\text{int}$, where
\begin{equation}
    T_\text{int} = N_\text{frame}T_\text{frame}
\end{equation}
with $N_\text{frame}$ the number of frames that are combined into a single exposure. The expression for $T_\text{frame}$ from \citet{stark2019} is
\begin{equation}
    T_\text{frame} = \mathfrak{Re}\left(-\frac{1}{\text{CR}_\text{peak}}\left(1+W_{-1}\left(-\frac{q}{e}\right)\right)\right),
\end{equation}
where $\text{CR}_\text{peak}$ is the count rate of the brightest pixel in the image, $W_{-1}$ denotes the lower branch of the Lambert $W$-function, $q$ is the probability that less than two photons arrive at the detector within $T_\text{frame}$ (also known as Geiger efficiency), and $e$ is the Euler number. We here adopt a Geiger efficiency of 99\% which we find to be a reasonable trade-off between avoiding coincidence losses while keeping the computation time short.

We consider images affected by either photon noise only or both photon and detector noise. In the former case, we simply simulate $N_\text{frame}$ individual frames with frame times of $T_\text{frame}$ by drawing the number of detected photons from a Poisson distribution with mean (and therefore variance) $\text{CR}_\text{image}T_\text{frame}$, where $\text{CR}_\text{image}$ is the count rate of the pixels in the noiseless images obtained according to the previous Section. Then, we stack the $N_\text{frame}$ individual frames together to obtain an exposure with integration time $T_\text{int}$. Figure~\ref{fig:exovista} shows a noiseless coronagraphic scene image on the left and the same image with photon noise added on the right. In the latter case, we use the \texttt{emccd\_detect} package \citep{nemati2020} with the detector parameters given in Table~\ref{tab:detector_parameters} to simulate $N_\text{frame}$ individual frames with both photon and detector noise. To speed up the computations, we set the read noise to 0~e-/frame because it is negligible given the high electron-multiplying (EM) gain. We also set the cosmic ray rate to 0~hits/cm$^2$/s because modeling cosmic rays is very time-consuming. While the \texttt{pc} (photon counting) branch of the \texttt{emccd\_detect} package comes with a \texttt{get\_count\_rate} function that enables counting the number of detected photons while also correcting for coincidence and thresholding losses according to \citep{nemati2020}, this correction destroys the Poisson-like distribution of the detected photons \citep{ludwick2022}. Since this would prevent us from theoretically predicting the expected noise, we here use the \texttt{get\_counts\_uncorrected} function from the \texttt{pc} branch of the \texttt{emccd\_detect} package that does count the number of detected photons without correcting for coincidence and thresholding losses (and thus conserving the Poisson-like distribution). The theoretically expected noise can then be obtained by multiplying the photon noise with the coincidence and thresholding loss factors according to Equation~8 of \citet{ludwick2022}, that is
\begin{equation}
\left(\frac{1-\exp(-\lambda)}{\lambda}\right)^{3/2}\left(\exp(-\tau/g)\right)^{1/2},
\end{equation}
where $\lambda$ is the expected number of detected photons per frame, $\tau$ is the counting threshold, and $g$ is the detector gain. To minimize thresholding losses, we use a detection threshold of 10\% of the EM gain.

\subsection{PSF and disk subtraction methods}
\label{sec:psf_and_disk_subtraction_methods}

\begin{figure*}
\includegraphics[width=\textwidth]{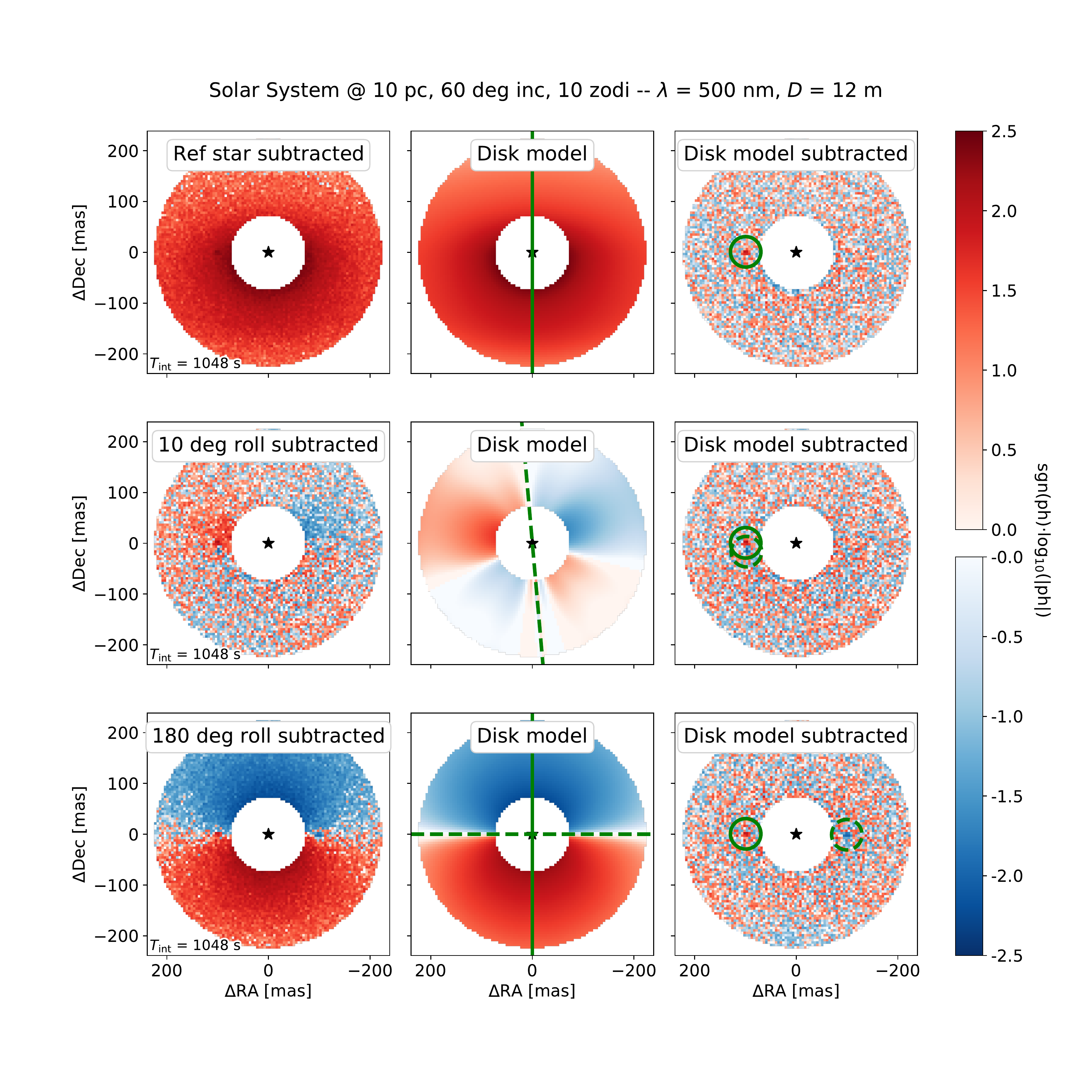}
\caption{Simulated coronagraphic images of the Solar System with 10~zodi of warm dust, at 60~deg inclination, and at a distance of 10~pc together with the best fit parametric disk models. Red represents positive and blue represents negative counts, both shown in a logarithmic color stretch. The reference star or roll subtracted images are shown in the left column, the best fit parametric disk models are shown in the middle column, and the residuals are shown in the right column. The top row shows the reference star subtraction, the middle row shows the 10~deg roll subtraction, and the bottom row shows the 180~deg roll subtraction case. The symmetry (solid green lines) and antisymmetry (dashed green lines) axes of the parametric disk models are highlighted in the middle column and the locations of the Earth twin are highlighted by solid green (science image) and dashed green (roll image) circles in the right column. The integration time is chosen to detect the Earth twin at a photon noise-limited SNR of 7 in the roll subtraction case and equals 1048~s.}
\label{fig:parametric_disk_model}
\end{figure*}

The main goal of this work is to study exozodi subtraction methods in the context of PSF subtraction in the $1\mathrm{e}{-10}$ contrast regime. This will help to develop an understanding of how well these background sources, especially the exozodi, can be mitigated for observations at contrasts sufficient to directly detect Earth-like planets in the habitable zone of Sun-like stars. The question is whether we will be able to use simple models to fit and subtract exozodi down to the photon noise limit?

\subsubsection{Calibration of stellar speckles}
\label{sec:calibration_of_stellar_speckles}

Our approach here is to simulate either an (ideal) reference star or a roll subtraction to calibrate the stellar speckles and the exozodi followed by fitting and subtracting a simple toy model of the residual flux from the exozodiacal dust disk. The reference star and the second roll observation are simulated using the second set of on-axis PSFs resembling a different speckle field and the reference star is chosen to be a copy of the science target but without exozodiacal disk or planets. Then, we will measure the noise in the residual images and compare it to the theoretical photon and detector noise limit. We choose these methods because reference star and roll subtraction are common observing modes for other space telescopes such as the \emph{James Webb Space Telescope} (\emph{JWST}). We note that in the $1\mathrm{e}{-10}$ contrast regime, reference star subtraction may not be feasible, as most systems are expected to harbor some level of exozodiacal dust. However, we include ideal reference star subtraction where the reference star has exactly the same properties (e.g., spectral type, angular diameter) as the science target as an informative limiting scenario in which the stellar speckles are measured and subtracted as precisely as possible. For the roll subtraction, we consider roll angles of 10~deg for LUVOIR-A and 20~deg for LUVOIR-B as well as 180~deg. The smaller roll angle is motivated by the position angle range accessible to \emph{JWST} and the requirement that the PSFs of an Earth twin at quadrature at a distance of 10~pc are separated by at least $2~\lambda/D$. The larger roll angle of 180~deg is expected to result in the highest possible symmetry in the roll-subtracted images. We note that enabling continuous access to such a large range of position angles for a significant fraction of the sky would require a different instrumental design compared to \emph{JWST} and studying the feasibility of such designs is beyond the scope of this work. Figure~\ref{fig:psf_and_disk_subtraction_methods} summarizes these methods defining REF as the PSF reference observation and SCI 1 and SCI 2 as the two science observations at different roll angles. To summarize, the three methods applied here are:
\begin{enumerate}
    \item subtracting REF from SCI 1 (ref subtraction) and calibrating with a parametric disk model (disk subtraction),
    \item subtracting SCI 2 from SCI 1 with a 10/20~deg roll angle (roll subtraction) and calibrating with a parametric disk model (antisymmetric subtraction),
    \item subtracting SCI 2 from SCI 1 with a 180~deg roll angle (roll subtraction) and calibrating with a parametric disk model (antisymmetric subtraction).
\end{enumerate}
We apply these data reduction methods to simulated images of 30 Solar System analogs at a distance of 10~pc with varying exozodi levels (1, 2, 5, 10, 20, 50, 100, 200, 500, 1000~zodi) and inclinations (0, 30, 60~deg).

\subsubsection{Parametric disk model}
\label{sec:parametric_disk_model}

The parametric disk model that we use to fit the residual exozodiacal disk flux after the reference star and roll subtraction is a polynomial in radius $r$ and either scattering angle or azimuth angle $\theta$ of the following form:
\begin{align}
    p(r,\theta) = \sum_{m=1}^{5}c_mr^m+\sum_{n=0}^{10}c_n\theta^n,
\end{align}
where the $c_m$ and $c_n$ are constant coefficients whose values are obtained by fitting the model to the images. We evaluate the radius and scattering angle in a single plane defined by the disk's inclination $i$ and position angle $\text{PA}_\text{disk}$, effectively fitting the disk as if it were infinitely thin. Furthermore, the scattering angle is defined as the angle between the host star, the scattering location, and the observer and the azimuth angle is defined as the angle between the host star, the scattering location, and the $\text{PA}_\text{disk}$-axis. In the case of the reference star subtraction, the calibrated images contain residual stellar speckles and the entire exozodiacal disk flux, assuming that the reference star has no exozodiacal disk itself. We assume that the exozodiacal disk flux is distributed symmetrically along the $\text{PA}_\text{disk}$-axis and put this as an additional constraint on our model that we evaluate for the scattering angle of the disk particles. This means that our model cannot fit asymmetric disk structures, but such features are also not included in our simulations. In case of the roll subtraction, the calibrated images also contain residual stellar speckles, but the residual exozodiacal disk flux is now distributed antisymmetrically along the $(\text{PA}_\text{roll 1}+\text{PA}_\text{roll 2})/2$-axis. We put this as an additional constraint on our model that we evaluate for the azimuth angle of the disk particles. Finally, in the special case of the 180~deg roll subtraction, the residual exozodiacal disk flux is both antisymmetric along the $(\text{PA}_\text{roll 1}+\text{PA}_\text{roll 2})/2$-axis and also symmetric along the $\text{PA}_\text{roll 1}$-axis = $\text{PA}_\text{roll 2}$-axis and we put both conditions as additional constraints on our model. Figure~\ref{fig:parametric_disk_model} shows the calibrated images next to the best fit parametric disk models and the residuals for all three cases and illustrates their symmetry properties.

\section{Results}
\label{sec:results}

\begin{figure*}
\centering
\includegraphics[width=0.8\textwidth]{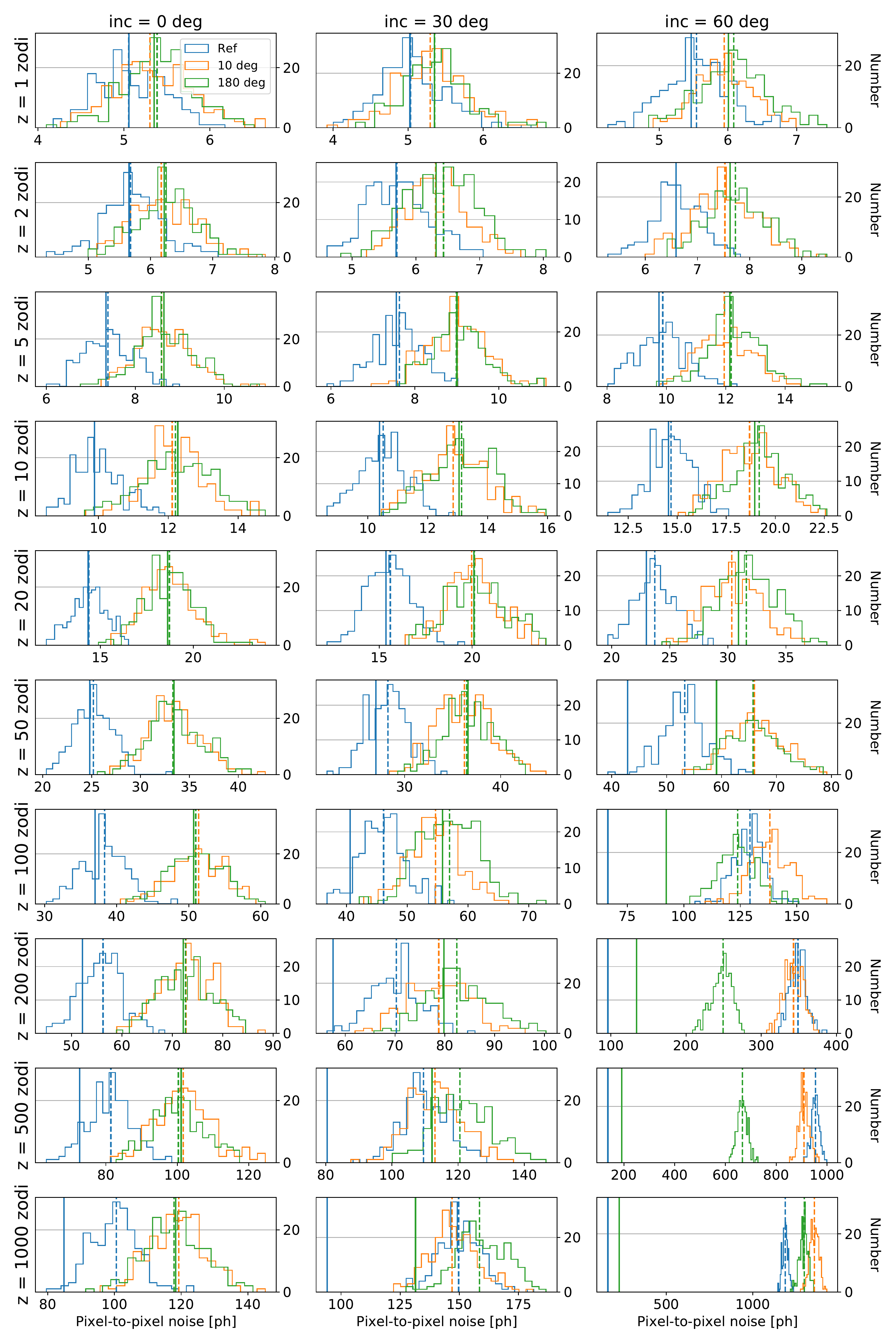}
\caption{Histograms of the measured pixel-to-pixel noise distribution in the 200 simulated observations of Solar System twins with inclinations of 0~deg (left column), 30~deg (middle column), and 60~deg (right column), exozodi levels from 1 to 1000~zodi (top to bottom row), and at a distance of 10~pc. The blue histograms show the reference star subtraction case and the orange and green histograms show the 10 and 180~deg roll subtraction cases. The solid and dashed vertical lines show the theoretically expected and the mean of the empirically measured noise, respectively.}
\label{fig:fix_inc_fit}
\end{figure*}

\begin{figure*}
\centering
\includegraphics[width=0.9\textwidth]{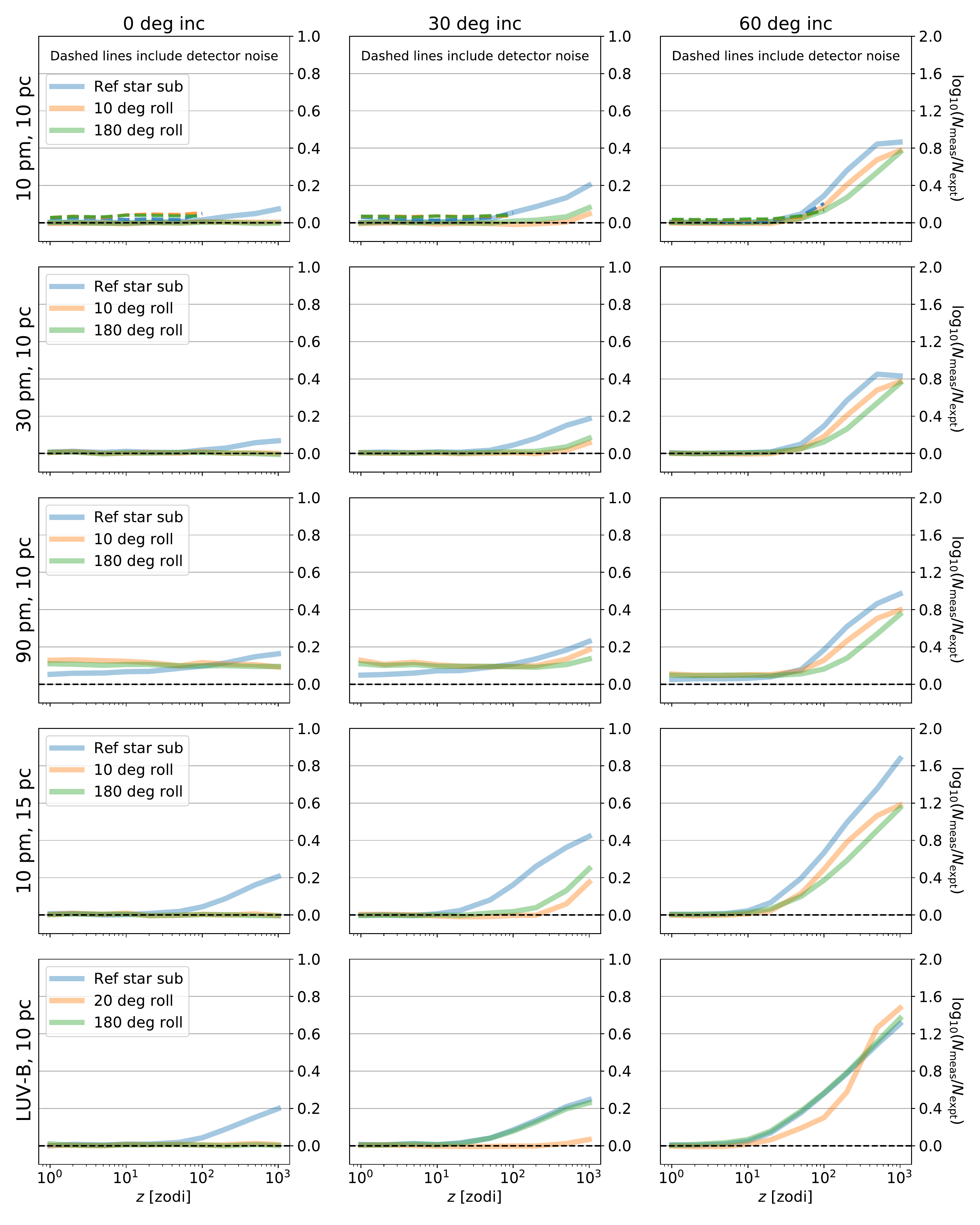}
\caption{Ratio of the measured and the theoretically expected noise as a function of the exozodi level for the 10~pm RMS WFE, 10~pc case (top row), the 30~pm RMS WFE, 10~pc case (second row), the 90~pm RMS WFE, 10~pc case (third row), the 10~pm RMS WFE, 15~pc case (fourth row), and the LUVOIR-B, 10~pc case (bottom row). Shown is the average over 200 simulated observations of Solar System twins with inclinations of 0~deg (left column), 30~deg (middle column), and 60~deg (right column). The blue curve shows the reference star subtraction case, the orange curve shows the 10/20~deg roll subtraction case, and the green curve shows the 180~deg roll subtraction case. In the top row, the dashed lines show the simulations including detector noise which were only computed up to 100~zodi due to exceedingly long computation times.}
\label{fig:noise_ratio}
\end{figure*}

We use the simulation tools described in the previous Section to simulate realistic observations of exoplanetary systems with a future reflected-light coronagraphy space telescope. Then, we study our ability to subtract the exozodi down to the theoretical noise limit, which has important implications for detecting Earth-like habitable zone planets at contrasts of $\sim1\mathrm{e}{-10}$. Therefore, we fit a simple toy model for the exozodiacal dust disk to the simulated coronagraphic images and measure the residual noise after subtracting the model. While this toy model is fit to the images globally (see Figure~\ref{fig:parametric_disk_model}), the residual noise is measured locally at the position of an Earth twin at quadrature\footnote{We note that there are two quadrature positions, one on each side of the star.}. When measuring the residual noise, we use a generously large photometric aperture with a radius of $3~\lambda/D$ in order to be sensitive to residual stellar speckles and uncalibrated exozodiacal disk flux around the planet location which could potentially cause confusion with a true point source. We repeat this procedure 200 times for each of the 30 different scenarios that we are studying here to obtain a robust statistical average. We ensure robustness by verifying that the mean value of the measured residual noise changes by no more than $1\%$ when doubling the sample size. We note that for the (ideal) reference star subtraction case, the theoretically expected noise $\text{CR}_\text{noise}$ is given as
\begin{equation}
\text{CR}_\text{noise} = 2\text{CR}_\text{star}+2\text{CR}_\text{det}+\text{CR}_\text{disk}
\end{equation}
since the exozodiacal dust disk is only present in the science but not in the reference star images. However, for the roll subtraction case, we have
\begin{equation}
\text{CR}_\text{noise} = 2\cdot(\text{CR}_\text{star}+\text{CR}_\text{det}+\text{CR}_\text{disk}).
\end{equation}
This means that the theoretical noise limit for the roll subtraction cases is larger than that for the reference star subtraction case. It should be noted, though, that the signal of a potential companion would also be larger for the roll subtraction cases since there would be a positive planet and a negative anti-planet in the roll-subtracted images. We emphasize that for better comparability, the integration time is the same in the reference star and the roll subtraction cases and is always based on Equation~\ref{eqn:tint} with the photon noise from the stellar speckle field and the exozodiacal disk as well as the detector noise as background and thus reflects an expected $\text{SNR} = 7$ detection of the Earth twin at quadrature in the roll subtraction case. Finally, the residual noise that we measure in the disk subtracted images is given as
\begin{equation}
    \sigma_\text{measured} = \sqrt{\sigma_\text{expected}^2+\sigma_\text{systematic}^2}
\end{equation}
and consists of an expected contribution from photon and detector noise and a systematic contribution of fit residuals after subtracting the best fit exozodiacal dust disk model from the calibrated images. Therefore, if our parametric disk model fits the simulated exozodi perfectly, we will only be left with the expected photon and detector noise. However, if our parametric disk model fails to reproduce the simulated exozodi perfectly, there will be additional noise from fit residuals that we will be able to measure in the disk-subtracted images.

For our baseline case, we keep the inclination of the parametric disk model for the reference star subtraction fixed at the true inclination of the system, assuming that the system inclination can be constrained from previous observations or using more sophisticated models for the exozodi. We also conduct a second set of simulations with the inclination as a free parameter in Section~\ref{sec:retrieving_the_inclination}. Our results are shown in Figure~\ref{fig:fix_inc_fit}. For each of the 30 different scenarios, we plot a histogram of the measured pixel-to-pixel noise in each of the 200 simulations for all three data reduction methods (reference star subtraction, 10/20~deg roll subtraction, 180~deg roll subtraction). Overlaid with dashed and solid lines are the means of the measured distributions and the theoretically expected photon noise limits, respectively.

For small exozodi levels of $\lesssim20~\text{zodi}$, our parametric disk model is able to subtract the exozodi down to the photon noise limit for any inclination and data reduction method. At $\sim50~\text{zodi}$, the model starts failing to completely subtract the exozodi for disks with $60~\text{deg}$ inclination. The same happens at $\sim100~\text{zodi}$ for disks with $30~\text{deg}$ inclination, although we note that the two roll subtraction methods still perform well up to $\sim500~\text{zodi}$ of dust. Unsurprisingly, due to the symmetry of the 0~deg inclination case, the two roll subtraction methods perform perfectly well for arbitrarily high exozodiacal dust levels. However, this case nicely illustrates that our parametric disk model is not complex enough to model disks with $\gtrsim500~\text{zodi}$ with the reference star subtraction method as the \texttt{exoVista} disk model features larger radial and azimuthal gradients due to grain-grain collisions within the disk than our simple toy model can reproduce.

A better visualization of the performance of the different data reduction methods as a function of the exozodi level is shown in Figure~\ref{fig:noise_ratio}. It confirms the previous statements and suggests that up to at least $\sim20~\text{zodi}$, the exozodi can be subtracted down to the photon noise limit even with a simple toy model for the exozodiacal dust disk. This is an encouraging finding given that the median exozodi level for Sun-like stars has been found to be $\sim3~\text{zodi}$ \citep[95\% confidence upper limit $\sim27~\text{zodi}$,][]{ertel2018,ertel2020}. In Figure~\ref{fig:noise_ratio}, it also becomes evident that while the 10~deg roll subtraction method performs slightly better than the 180~deg one in the 30~deg inclination case, it is the other way around in the 60~deg inclination case. This is because in the 30~deg inclination case, the exozodiacal dust disk appears smoother than in the 60~deg inclination case, so that a small roll angle leads to more similar regions of the disk being subtracted from one another, resulting in smaller residuals.

\section{Discussion}
\label{sec:discussion}

\subsection{Impact of wavefront errors}
\label{sec:impact_of_wavefront_errors}

For the baseline case, we used a coronagraph model with an optimistic wavefront drift of 10~pm RMS between the science and the reference/roll target observations. As can be seen in Figure~\ref{fig:aplc}, the noise floor of the residual speckles at the separation of an Earth twin at a distance of 10~pc (which is $\sim100$~mas or $\sim11.5~\lambda/D$) is almost $1\mathrm{e}{-12}$ and thus about two orders of magnitude fainter than the Earth twin itself. The 1~zodi face-on Solar System dust disk is of similar brightness as the Earth twin so that the residual stellar speckles have virtually no impact on the performance of our toy model to fit and subtract the exozodi. Hence, in this Section, we repeat the previous simulations with more conservative wavefront drifts of 30~pm RMS and 90~pm RMS. This leads to a brighter noise floor of residual speckles in the reference star- and roll-subtracted images. As can be seen in Figure~\ref{fig:aplc}, larger wavefront drifts also impact the raw APLC contrast. Consequently, the integration times to detect an Earth twin at quadrature at an SNR of 7 are longer than in the previous simulations with a wavefront drift of only 10~pm RMS. 

The second and third rows of Figure~\ref{fig:noise_ratio} show the exozodi subtraction results for the 30~pm RMS and the 90~pm RMS cases. For the 30~pm RMS case, no significant difference can be observed with respect to the baseline case with 10~pm RMS. This could be expected given that the noise floor of the residual stellar speckles is still fainter than the Earth twin and the 1~zodi dust disk in this case. However, a significant difference especially for the cases with a low exozodi level can be observed for the 90~pm RMS case. The noise floor of the residual stellar speckles is now on the order of (or even brighter than) the exozodi at least for exozodi levels below $\sim10$~zodi. The brightness of the residual stellar speckles is above the photon (i.e., Poisson) noise expected from the raw APLC contrast, adding a systematic noise floor of $\sim1.1$--1.3 times the expected photon noise. While the measured noise is dominated by the residual stellar speckles for exozodi levels below $\sim10$~zodi and should therefore be similar for the reference star and the roll subtraction case, the expected photon noise is still computed from the raw APLC contrast and therefore dominated by the exozodiacal disk flux. Hence, the expected photon noise is higher in the roll subtraction case (since the exozodi contributes twice there, in both the science and the roll image), resulting in a longer integration time. Consequently, the roll subtraction cases appear to perform worse than the reference star subtraction case for low exozodi levels. However, once the exozodiacal disk flux becomes similarly bright or brighter than the residual stellar speckles, the performance of our toy model to fit and subtract the exozodi is comparable to the baseline case with 10~pm RMS WFE. We therefore conclude that while residual stellar speckles can introduce significant systematic noise to the observations and potentially mimic the signal of planets, high wavefront errors have no impact on the performance of our toy model to subtract the exozodi because the exozodi is a smooth and extended source that can be fit by our toy model as soon as it is the dominant noise term. We note that this conclusion might not hold for strongly inclined or highly structured exozodiacal dust disks whose reflected light might be confused with residual stellar speckles more easily.

\subsection{Impact of distance}
\label{sec:impact_of_distance}

For a future exo-Earth searching space mission, the exozodiacal dust disk can usually be assumed to be resolved, meaning that its surface brightness will be independent of the distance to the observed exoplanetary system, whereas the planet and star become fainter by a factor of one over the distance squared. Therefore, the exozodi of a further away exoplanetary system will have a larger contribution to the total astrophysical noise. To investigate this effect, we simulate a set of observations of an F-type star at a distance of 15~pc. The target was chosen so that its habitable zone and its Earth analog orbit appear at the same separation of 100~mas as in the previous simulations of a Solar System twin at a distance of 10~pc. The reason for this choice is that the 18\% bandwidth medium IWA coronagraph has an inner working angle of $\sim75$~mas and we need to be able to place the entire $3~\lambda/D$ photometric aperture around the quadrature position of an Earth twin outside of this IWA. Hence, for the F-type star at 15~pc, we scale the orbital distances of the planets and the exozodiacal dust components by $\sqrt{L/L_\odot} = 1.5$, where $L$ is the luminosity of the F-type star and $L_\odot$ is the luminosity of the Sun.

The fourth row of Figure~\ref{fig:noise_ratio} shows the exozodi subtraction results for the 15~pc case and reveals that the measured noise deviates from the theoretically expected photon noise already at lower exozodi levels as for the 10~pc case. This can be explained by the total integration time now being longer (since a potential Earth twin would be fainter at a larger distance) and thus more exozodiacal disk flux being collected and needing to be subtracted. The maximum exozodi level at which the exozodi can still be subtracted down to the photon noise limit decreases from $\sim50$~zodi in the baseline case to $\sim10$~zodi in the 15~pc case. For higher exozodi levels, yield studies should include a systematic noise term on top of the photon noise from the exozodi to penalize the optimization routine for observing such systems.

\subsection{Impact of detector noise}
\label{sec:impact_of_detector_noise}

So far, we have only considered simulations affected by photon (i.e., Poisson) noise. However, realistic observations will also be affected by detector noise such as dark current and clock-induced charge. To simulate such detector noise for a detector operated in Geiger (i.e., photon-counting) mode, we use the \texttt{emccd\_detect} package\footnote{\url{https://github.com/wfirst-cgi/emccd_detect}} originally developed for the \emph{Roman} CGI instrument \citep{kasdin2020}. The package and the adopted detector parameters are described in Section~\ref{sec:photon_and_detector_noise} and Table~\ref{tab:detector_parameters}. Cosmic rays are not included in the simulations. Based on the results from Section~\ref{sec:impact_of_wavefront_errors}, we expect that localized systematic noise from cosmic rays should not affect the performance of our toy model to fit and subtract the exozodi, similar as it is the case for residual stellar speckles. While they might impact our ability to detect point-sources such as exoplanets, this will be the topic of another paper. Simulating a frame with detector noise is much slower than simulating one with only photon noise so that our simulations become very time-consuming for exozodi levels beyond 100~zodi. If the exozodi is bright, the frame time needs to be short to avoid coincidence losses (see Section~\ref{sec:photon_and_detector_noise}) while the total integration time must be long to detect an Earth twin at quadrature at an SNR of 7. Hence, on the order of 100'000 frames or more would need to be simulated for an observation of a system with an exozodi level beyond 100~zodi. We therefore decided to consider detector noise only for the baseline scenario with a wavefront error of 10~pm RMS and below 100~zodi.

The top row of Figure~\ref{fig:noise_ratio} shows the ratio of the measured noise and the theoretically expected photon and detector noise obtained from the simulations with detector noise (dashed lines), overlaid on the baseline scenario with only photon noise (solid lines). It can be seen that the additional detector noise with the noise parameters given in Table~\ref{tab:detector_parameters} has virtually no impact on the performance of our toy model to fit and subtract the exozodi. This is the case because the expected noise parameters are small and negligible over the photon noise from the residual stellar speckles and the exozodiacal dust.

\subsection{Impact of aperture size}
\label{sec:impact_of_aperture_size}

Our baseline case uses a 12~m circumscribed aperture APLC corresponding to a descoped version of the LUVOIR-A design presented in the LUVOIR Final Report \citep{luvoir2019} to create a high-contrast region between $\sim8.5$--$26~\lambda/D$ where small Earth-like exoplanets can be detected in reflected light. In accordance with the Astro2020 Decadal Survey on Astronomy \& Astrophysics, we also consider a smaller 6.7~m inscribed diameter (8~m circumscribed diameter) aperture VC6 coronagraph equivalent to the LUVOIR-B design presented in the LUVOIR Final Report which achieves the high-contrasts required to detect exo-Earths already beyond $\sim2.5~\lambda/D$. The decrease in angular resolution due to the smaller aperture is expected to have similar consequences as the increase in distance investigated in Section~\ref{sec:impact_of_distance}. We note that the distance increase by a factor of 15~pc/10~pc = 1.5 in Section~\ref{sec:impact_of_distance} is equivalent to the circumscribed aperture diameter decrease by a factor of 12~m/8~m = 1.5 in this Section so that any differences between the two cases can be traced back to differences in the underlying coronagraphic speckle maps (Figure~\ref{fig:speckles}) or the different sampling of the simulated scene (we note that the scene was scaled up by a factor of 1.5 in the 15~pc case so that it has the same apparent size as in the 10~pc case, but the LUVOIR-B pixel scale is coarser than the one of LUVOIR-A).

The bottom row of Figure~\ref{fig:noise_ratio} shows the ratio of the measured noise and the theoretically expected photon noise obtained from the simulations with this smaller 8~m circumscribed aperture and reveals a much earlier deviation from the photon noise limit already for exozodi levels $\gtrsim50$~zodi in the face-on case and $\gtrsim10$~zodi in the 60~deg inclined case. This behavior mirrors the one observed for a more distant system in Section~\ref{sec:impact_of_distance} and means that a flagship mission with a smaller 8~m circumscribed aperture would be \sout{significantly} more affected by the contamination from exozodiacal dust. While the 20~deg roll subtraction case performs significantly better than the reference star and the 180~deg roll subtraction cases for inclined systems (at least for moderate exozodi levels), it has to be noted that this stems from a significant amount of disk self-subtraction near the inner edge of the large $3~\lambda/D$ photometric aperture used to empirically measure the pixel-to-pixel noise in the images. Furthermore, the reference star subtraction case with LUVOIR-B performs slightly better than with LUVOIR-A at a distance of 15~pc because the scene is sampled more coarsely in the former case resulting in a smoother exozodi that can be subtracted more easily with our parametric disk model. Nevertheless, while our parametric disk model is only a simple toy model for the exozodi, our findings show that calibrating the exozodi down to the photon noise limit is more challenging for a mission with a smaller 8~m circumscribed aperture and needs to be investigated more carefully in future yield studies.

\subsection{Retrieving the inclination}
\label{sec:retrieving_the_inclination}

\begin{figure*}
\centering
\includegraphics[trim={0 4cm 0 4cm},clip,width=\textwidth]{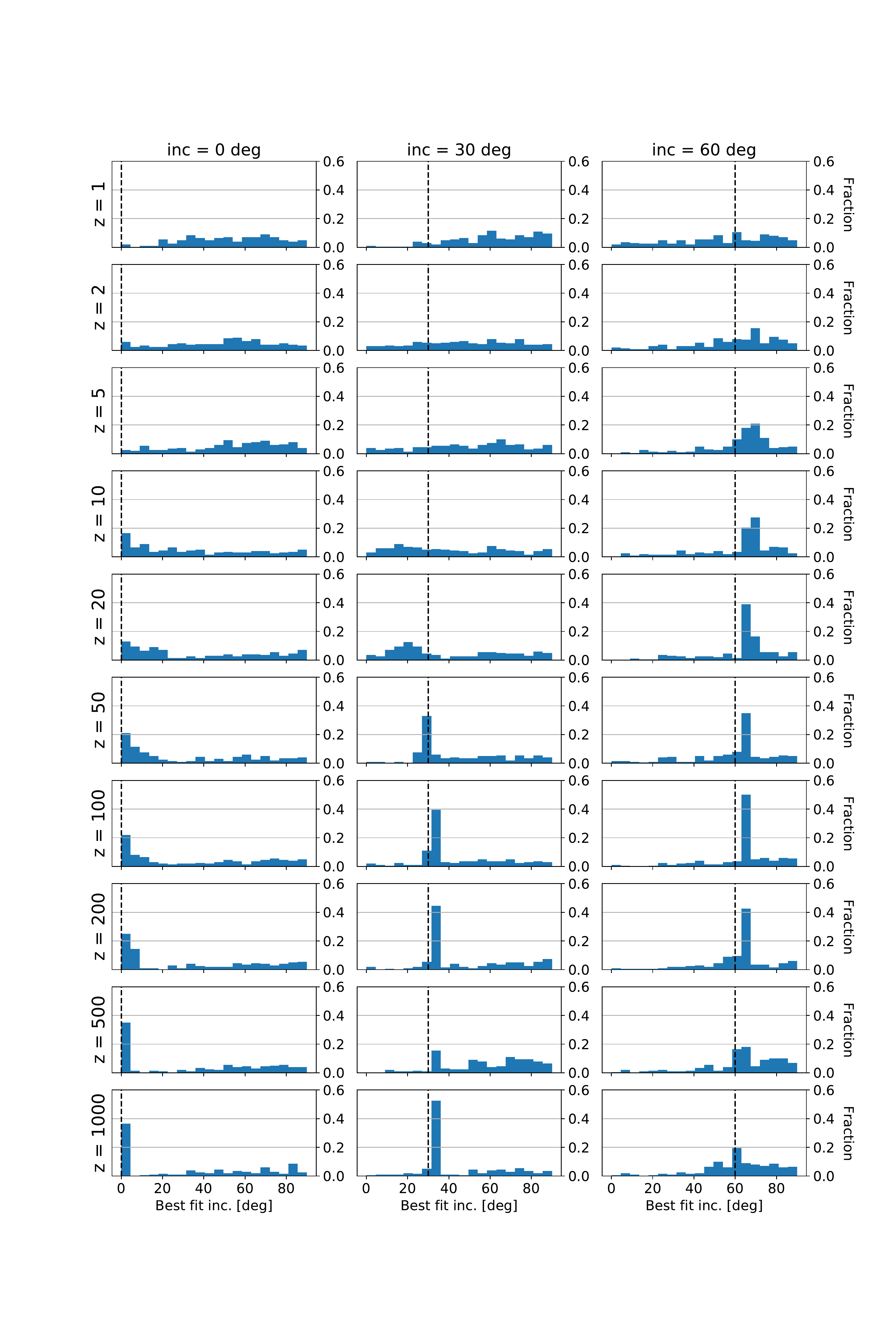}
\caption{Distribution of the best fit inclination retrieved from the fits with free inclination. The true system inclination is highlighted by a dashed black line. For each scenario, 200 individual fits were performed with the prior of the inclination $i$ being randomly distributed so that $\cos(i) \in [-1, 1)$.}
\label{fig:free_inc_fit}
\end{figure*}

In all previous simulations, we have always assumed that the inclination of the exozodiacal dust disk is known \textit{a priori} and evaluated our parametric disk model in the midplane of the astrophysical scene.\footnote{We note that this is only relevant for the reference star subtraction case, in which the disk model is evaluated in physical distance from the star and scattering angle. In the roll subtraction case, we are always fitting the antisymmetric residual from the disk self-subtraction in polar coordinates with zero inclination with respect to the image plane.} To investigate whether the disk inclination can be retrieved with our simple parametric disk model, we repeat the baseline reference star subtraction case with inclination as a free parameter, using a uniformly distributed prior between -1 and 1 in $\cos(i)$, where $i$ is the inclination. This assumes that the disk midplane is distributed uniformly on a sphere. Figure~\ref{fig:free_inc_fit} shows the distribution of the best fit inclination for the 30 different scenarios considered in the baseline case (Solar System at a distance of 10~pc and with a wavefront error of 10~pm RMS). As can be seen in the top few rows of the Figure, the distribution of the best fit inclination appears to be random for the scenarios with low exozodi levels. This results from the exozodi being composed of a rather smooth distribution of dust and the convolution with the coronagraph PSF further blurring the observed image and obscuring the true inclination of the disk. Once the exozodi level becomes high enough, though, the injected inclination is correctly retrieved in a majority of the fits, where we consider a best fit within $\pm10$~deg as correctly retrieved. The turnover point appears to be around $\sim200$~zodi for the 0~deg inclination case, around $\sim50$~zodi for the 30~deg inclination case, and around $\sim20$~zodi for the 60~deg inclination case. We explain the shift of the turnover point towards smaller exozodi levels for increasing system inclination with stronger scattering features becoming visible at higher inclination and aiding the fit with retrieving the correct inclination. For the 60~deg inclination scenario, the retrieved inclination starts to become more random again at exozodi levels $\gtrsim500$~zodi. At such high exozodi levels, a prominent dust ring becomes visible at high system inclination and our simple parametric disk model is not suited to reproduce the morphology of such a ring resulting in worse performance.

It is noteworthy that our parametric disk model describes a rather smooth exozodi and therefore produces no good fit for systems with high exozodiacal dust levels which tend to have a ring-like structure stemming from collisional interactions between the dust particles. The field-of-view of our simulated coronagraphic images is limited to about 200~mas from the host star, so that the bright ring-like feature of the exozodiacal dust disk only becomes visible at 60~deg inclination (and $z \gtrsim 500$). For these systems, it would be possible to retrieve the inclination of the exozodi by fitting a more sophisticated inclined ring model. In general, we note that more sophisticated disk models might be better suited for retrieving the inclination of the exozodi. However, firstly it would not make sense to consider an exozodi model that is as sophisticated as the injected one and secondly our main goal is to illustrate that retrieving the inclination of an exoplanetary system by measuring the inclination of the exozodi is not straightforward at all, at least for systems with low exozodi levels. This has direct consequences for an exo-Earth searching future direct imaging mission like the one recommended in the Astro2020 Decadal Survey on Astronomy \& Astrophysics since measuring the inclination via planet orbits might require at least three observations of a planet in the system over multiple epochs temporally separated by months. We note that inferring the inclination of an exoplanetary system from a single visit (i.e., from the exozodi) would be valuable, however, as it would enable deriving the orbital separation of potential planets in the system and understanding whether they lie within the habitable zone or not which in turn enables for faster decision making for the characterization phase of the mission.

\section{Summary and conclusions}
\label{sec:summary_and_conclusions}

In this paper, we aim to address the question whether the contamination from exozodiacal dust in high-contrast reflected-light coronagraphy observations can be subtracted down to the photon noise limit. This assumption has readily been made for simplicity in previous yield estimates for future exo-Earth searching NASA and ESA flagship missions \citep{stark2014,stark2019,morgan2019,quanz2021} without further justification. Studying the validity of this assumption is critical for the correct assessment of the potential exo-Earth candidate yield of such future missions and to estimate their scientific capabilities \citep{roberge2012}.

To achieve this goal, we first develop a fast coronagraphic imaging simulator that is able to convolve an astrophysical scene from the \texttt{exoVista} tool \citep{stark2022} with a library of position-dependent coronagraphic PSFs within less than a second per image. Then, we generate simulated observations of a synthetic Solar System located at a distance of 10~pc with different exozodi levels and inclinations. Using these simulations, we perform (ideal) reference star subtraction and roll subtraction followed by fitting a simple toy model for the residual exozodiacal disk flux and subtracting it from the calibrated images. Then, we empirically measure the pixel-to-pixel noise in the final images and compare it to the theoretically expected photon and detector noise in order to assess whether the exozodi can be calibrated down to the photon and detector noise limit. Finally, we also investigate the impact of larger wavefront errors, larger system distance, and smaller aperture size on the performance of our toy model to fit and subtract the exozodi.

Our simulations indicate that with a simple parametric disk model and no prior knowledge on a given system, the exozodi can be subtracted down to the photon noise limit for exozodi levels up to $\sim50~\text{zodi}$ in the baseline case. While larger wavefront errors introduce an additional noise floor from residual stellar speckles, they do not impact the performance of our toy model to fit and subtract the exozodi. Larger system distance and smaller aperture size (here 8~m circumscribed diameter instead of 12~m) have a significant (negative) impact and decrease the maximum exozodi level at which the exozodi can be calibrated down to the photon noise limit to $\sim10$~zodi at least for highly inclined systems. It is noteworthy that the penalty factor is hence negligible for all considered cases if the exozodi level is below $\sim10$~zodi which is more than the median expected exozodi level for Sun-like stars of 3~zodi inferred from the HOSTS survey \citep{ertel2020}.

Based on our simulations, we derive a penalty factor that can be used in future yield studies to account for the systematic excess noise in systems with a high exozodiacal dust level and inclination. This excess noise factor, shown in Figure~\ref{fig:noise_ratio}, accounts for the residual exozodiacal disk flux that cannot be subtracted down to the photon noise limit in cases with high exozodi levels and inclinations. The approach would be to first parametrize the penalty factor as a function of the exozodi level and system inclination. Then, when an observation of a given star is simulated with a yield estimate code, the corresponding penalty factor could be obtained and multiplied with the pure photon noise used in current yield estimate codes when determining whether an exoplanet is detectable or not. Of course, this is only possible if the yield study is simulated on a discrete sample of stars for which the exozodi level and system inclination is known a priori. When trying to determine the optimal integration time for each star in a sample one needs to consider each exoplanetary system as a set of random variables with a certain random distribution (e.g., the exozodi level, the system inclination, the number and properties of the putative planets in the system). One possibility for instance would then be to marginalize over all possible exozodi levels and system inclinations in a Monte-Carlo-like fashion and always use the corresponding penalty factor when computing the expected detection noise floor.

Furthermore, our toy model fits to high-fidelity simulated astrophysical scenes suggest that it will be difficult to use the exozodi to constrain the inclination of the exoplanetary systems from single-visit observations, at least as long as the exozodi level is below $\sim100$~zodi and no ring structure becomes evident in the images. This finding impacts the design of the search phase of a future exo-Earth searching direct imaging mission as multiple visits will be required to derive the exoplanet orbit and to ensure a successful deep follow-up observation in the characterization phase of the mission.

In future work, we will investigate matched filtering as an alternative method to maximize the SNR of an Earth-like companion in the presence of random and systematic errors from residual exozodiacal light. Preliminary experiments with our simulations have shown that this method might be promising for highly inclined and non-smooth exozodis. We will further use our simulation tool to increase the fidelity of yield studies for a future exo-Earth searching space mission. This will involve coupling our tool with the Altruistic Yield Optimization routine from \citet{stark2014,stark2015,stark2019} to model the realistic execution of an observing sequence and implement real-time decision making metrics and optimization routines.

\vspace{-0.5cm}

\begin{acknowledgments}

\section{Acknowledgments}
\label{sec:acknowledgments}

This work was performed by the Virtual Planetary Laboratory Team, a member of the NASA Nexus for Exoplanet System Science, funded via NASA Astrobiology Program Grant No. 80NSSC18K0829. The manuscript was substantially improved following helpful comments from an anonymous referee.

\end{acknowledgments}


\bibliography{sample631}{}
\bibliographystyle{aasjournal}

\begin{appendix}

\section{Additional noise histogram figures}

\begin{figure*}
\centering
\includegraphics[width=0.8\textwidth]{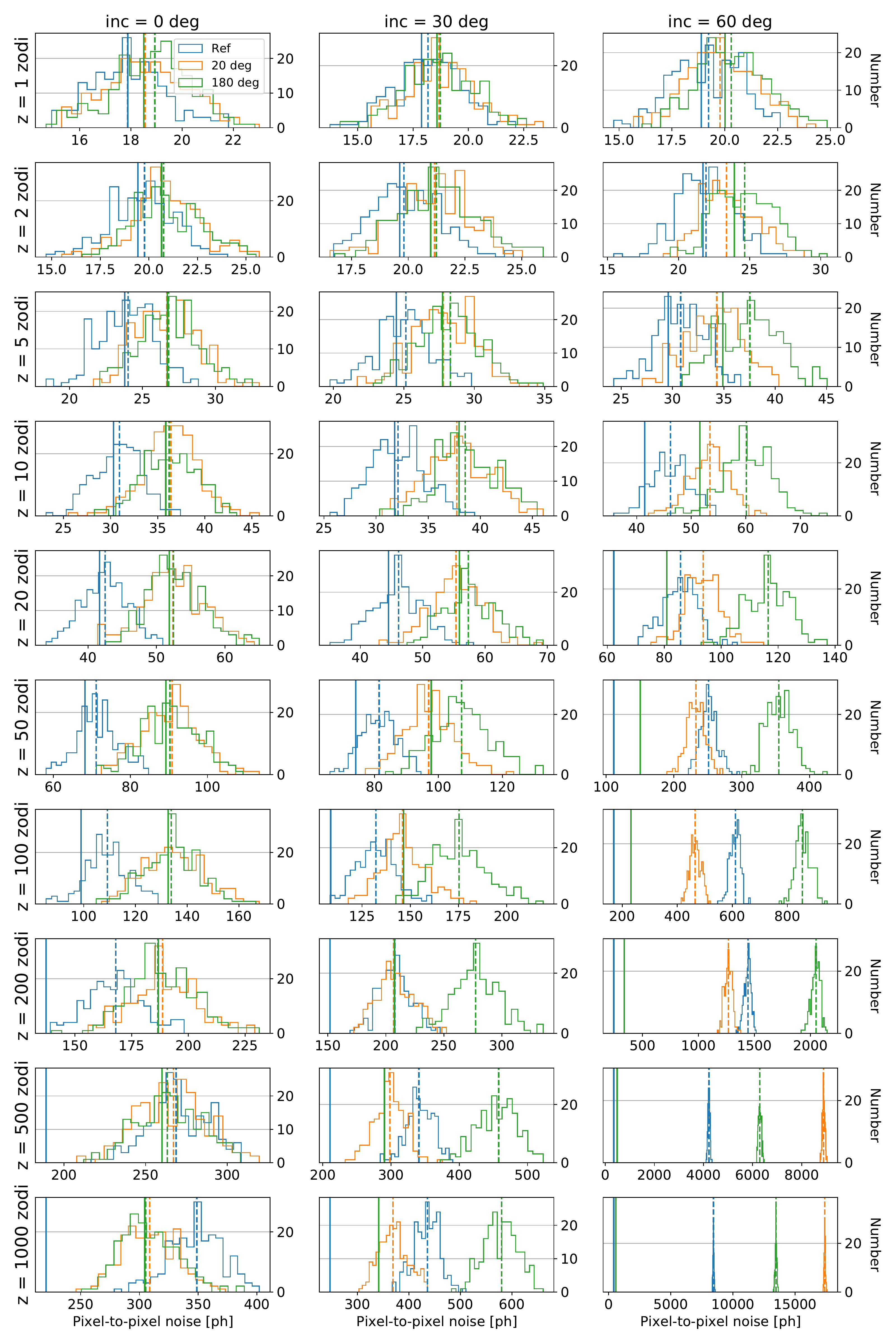}
\caption{Same as Figure~\ref{fig:fix_inc_fit}, but for the vector vortex charge 6 coronagraph used for the LUVOIR-B simulations.}
\label{fig:fix_inc_fit_luvb}
\end{figure*}

\end{appendix}



\end{document}